\documentclass[11pt]{article}
\usepackage[english]{babel}
\usepackage[utf8]{inputenc}
\usepackage[nottoc]{tocbibind}
\usepackage{graphicx}
\usepackage{amsmath}
\usepackage{amssymb}
\usepackage{amsxtra}

\def\institute#1{\gdef\@institute{#1}}

\title{The bound state of dark atom with the nucleus of substance}
\author{T.E. Bikbaev$^{1,2*}$, M.Yu. Khlopov$^{3}$, A.G. Mayorov$^{1}$\\
$^{1}$ National Research Nuclear University MEPhI \\115409 Moscow, Russia;\\
$^{2}$ Institute of Physics, Southern Federal University\\ Stachki 194 Rostov on Don 344090, Russia;\\
$^{3}$  Virtual Institute of Astroparticle physics,\\ 75018 Paris, France; e-mail: khlopov@apc.univ-paris7.fr \\ 
$^{*}$ Correspondence: bikbaev.98@bk.ru (T.B.)}

\date{November 2025}

\begin{document}
\maketitle

\begin{abstract}
The hypothesis of composite $X$He dark atoms offers a compelling framework to address the challenges in direct dark matter particles detection, as their neutral, atom-like configuration evades conventional experimental signatures. A critical issue may arise in interaction between $X$He and atomic nuclei due to the unshielded nuclear attraction, which could destabilize the dark atom’s bound state. To resolve this, we propose a novel numerical quantum mechanical approach that accounts for  self-consistent electromagnetic-nuclear couplings. This method addresses to eliminate the inherent complexity of the $X$He-nucleus three-body system, where analytical solutions are intractable. By reconstructing the effective interaction potential — including dipole Coulomb barrier and shallow potential well — we demonstrate how these features lead to the formation of $X$He-nucleus bound states and modulate low-energy capture processes. Our model enables validation of the dark atom hypothesis, particularly in interpreting experimental anomalies like annual modulation signals observed in DAMA/LIBRA. These findings advance the theoretical foundation for dark matter interactions and provide a robust framework for future experimental design.
\end{abstract}



\section{Dark atoms of $X$-helium}

The non-baryonic nature of dark matter necessitates the existence of novel, stable forms of non-relativistic matter. A particle-based origin for dark matter implies physics beyond the Standard Model, requiring new, stable fundamental particles. Among the candidates put forth are stable particles bearing electric charge \cite{KHLOPOV_2013, Bertone_2005, scott2011searches, belotsky2006composite, Belotsky2008}. This work examines the nuclear-interacting dark atom scenario, which reveals the composite essence of dark matter \cite{Kouv3, Kh_2008, Kh_2013, Beylin2020}. While the specific electric charge of new stable, multicharged particles is not predetermined a priori, rigorous experimental bounds significantly restrict the viable configurations, permitting solely stable, negatively charged states with a charge multiplicity of $-2n$ \cite{Cudell:2012fw, bulekov2017search}, where $n$ is a natural number. Herein, these particles are designated as $X$, with the particular instance of a doubly charged particle, $X$ with charge of $-2$, being referred to as $O^{--}$.

Thus, this work investigates composite dark matter model wherein hypothetical, stable, heavy, multicharged $X^{-2n}$ particles, possessing leptonic-like properties (absence or significant suppression of QCD interactions), combine with $n$ primary helium-4 nuclei to form electrically neutral, atom-like states through standard Coulomb attraction. These composite systems are designated as $X$He dark atoms. The $X^{-2n}$ particles may either be lepton-like in nature or arise from exotic, novel heavy quark families, characterized by weak-scale interaction cross-sections with standard model hadrons \cite{Khlopov_2020}.

The structural properties of bound dark atom system are governed by key parameter defined as $a \approx Z_{\alpha} Z_X \alpha A_{nHe} m_p R_{nHe}$. In this expression, $\alpha$ is the fine-structure constant, $Z_X$ and $Z_{\alpha}$ correspond to the charge numbers of the $X$ particle and the $n$He nucleus, $m_{p}$ denotes the proton mass, $A_{nHe}$ signifies the mass number of the $n$He nucleus, and $R_{nHe}$ represents its radius. Physically, the parameter $a$ quantifies the ratio of the dark atom's Bohr radius to the radius of the $n$-helium nucleus. The value of this ratio dictates the transition between two distinct structural regimes: Thomson-like configuration is realized when the Bohr radius of the $X$He atom is less than the $n$-helium nucleus radius, whereas Bohr-like atomic structure is adopted in the opposite case.

Within the parameter range $0 < a < 1$, the $X$He system conforms to the Bohr atom picture. In this regime, the helium nucleus can be treated as a point-like particle executing an orbital motion around a central, massive $X$ particle with negative charge. In the complementary domain $1 < a < \infty$, the system's structure is more accurately described by Thomson’s atomic model. Here, the helium nucleus, which can no longer be considered point-like, undergoes oscillatory motion around the heavier, negatively charged $X$ particle, resulting in a more diffuse and distributed atomic configuration.

The distinct properties of dark atoms lead to a scenario of "warmer-than-cold dark matter" during the formation of large-scale cosmic structure. While this model necessitates further detailed study, its predictions are consistent with the precision cosmological data \cite{Khlopov_2020}. The timeliness and significance of this research are driven by the necessity to deepen the investigation into the nuclear interaction characteristics of dark atoms and to evaluate the potential influence of $X$-helium on processes of nuclear transformations. A comprehensive understanding of these interactions is paramount for accurately assessing the contribution of dark atoms to primordial nucleosynthesis, the evolution of stars, and a spectrum of other physical, astrophysical, and cosmological phenomena in the early Universe \cite{Khlopov:2010ik}.

\section{Low energy binding of dark atoms to nuclei}

Direct dark matter detection experiments is characterized by a diversity of outcomes, underscoring the complex nature of potential interactions between dark matter candidates and subterranean detector materials. The $X$-helium model offers a potential resolution to the apparent conflicts among different direct dark matter detection experiments, which may stem from the unique mechanisms through which dark atoms engage with ordinary matter. Instances of this contradiction include the positive signals registered by the $DAMA/NaI$ and $DAMA/LIBRA$ experiments, which appear inconsistent with the null results reported by experiments including $XENON100$, $LUX$, and $CDMS$, among others \cite{BERNABEI_2020}.

The slowing down of cosmic $X$He within the terrestrial crust precludes the application of conventional direct detection techniques, which rely on identifying nuclear recoil signatures from Weakly Interacting Massive Particles (WIMPs) colliding with target nuclei. Nevertheless, the collisions of slow-moving $X$-helium atoms with these nuclei can result in the formation of low-energy bound states, a process described by the reaction:

\begin{equation}
A + (^4He^{++}X^{--}) \rightarrow [A(^4He^{++}X^{--})] + \gamma.
\end{equation}

It is assumed that within the uncertainties inherent to nuclear parameters, a certain range exists where the binding energy for the $X$He--$Na$ system falls within the 2--4 keV range \cite{Khlopov_2020}, representing a comparatively subtle energy scale. The capturing of dark atoms into such a bound state results in the deposition of an equivalent energy quantum, detectable as an ionization signal in NaI(Tl) crystal detectors like those used in the DAMA experiment. The resultant concentration of $X$He within the materials of underground detectors is governed by a balance between the falling cosmic flux of dark atoms and their diffusive transport towards the Earth's center. The availability of $X$-helium in the terrestrial subsurface is rapidly regulated by the kinematics of dark atom interactions with ordinary matter; it closely tracks variations in the incoming cosmic $X$He flux. Consequently, the capture rate of dark atoms is expected to exhibit annual modulation, which should be directly reflected as a corresponding periodic variation in the ionization signal originating from these capture events.

A direct result of the proposed model is the emergence of anomalous superheavy isotopes of sodium within the NaI(Tl) detector material of the DAMA experiment. The mass of these anomalous isotopes exceeds that of standard sodium isotopes by approximately the mass of the $X$ particle \cite{Khlopov:2010ik}. In contrast, the formation of analogous superheavy isotopes of iodine and thallium is improbable, as the formation of bound states between dark atoms and these nuclei is unfavorable \cite{Khlopov:2010ik}. Should these anomalous sodium atoms remain in a non-fully ionized state, their transport properties are governed by atomic cross-sections, resulting in a mobility reduced by approximately nine orders of magnitude compared to that of $O$He \cite{Khlopov:2010ik}. This reduction in mobility effectively leads to their accumulation and retention within the detector material. Consequently, mass spectroscopic examination of this substance could offer a crucial independent test for verifying the $X$-helium origin of the DAMA signal. Any methodology employed for such an analysis must account for the fragile nature of the $X$He-$Na$ bound systems, characterized by binding energies of merely several keV \cite{Beylin:2019gtw}.

The anticipated energy release in detector materials alternative to $NaI$ is predicted to occupy a spectral range predominantly above 2--6 keV \cite{Khlopov:2010ik}. Furthermore, such a signal may be entirely absent in detectors utilizing heavy target nuclei, such as xenon \cite{Khlopov:2010ik}.

\begin{figure}[h!]
\centering
\includegraphics[scale=0.8]{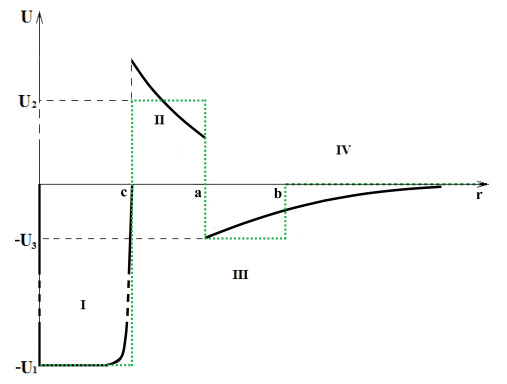}
\caption{
Hypothetical qualitative image of the shape of the effective interaction potential of $X$He dark atom with the nucleus of atom of matter \cite{Khlopov:2010ik}.}
\label{fig:Effective_potential}
\end{figure}

The unscreened nuclear charge of the dark atom introduces the possibility of a strong interaction between $X$He and terrestrial nuclei. This interaction can dissociate the dark atom's bound system, potentially giving rise to anomalous isotopes whose abundance in the environment is stringently constrained by existing experimental data \cite{Cudell:2012fw}. To resolve this challenge, the $X$He hypothesis postulates the existence of potential well in conjunction with a repulsive potential barrier within the effective interaction potential (see Figure \ref{fig:Effective_potential}). This potential structure inhibits the merger of the dark atom's constituents — namely, the $n$--He nucleus and the $X$ particle — with nuclei of ordinary matter. This specific feature of the interaction potential constitutes a critical prerequisite for the phenomenological viability of the $X$-helium hypothesis.

The specific profile of this effective potential (see Figure \ref{fig:Effective_potential}) arises principally from the rivalry between the electromagnetic repulsion and the attractive nuclear force originating from the nuclear shell of the dark atom and the target nucleus.

The total effective interaction potential of the $X$He--nucleus system is defined as the aggregate potential experienced by the nucleus of matter under the influence of the various forces emanating from the dark atom. This description holds within a coordinate system centered on the dark atom, when the nucleus undergoes a slow approach towards the dark atom from an initial distance much large compared to the characteristic sizes of the particles.

Eventually, the capture mechanism proceeds as follows: nucleus, moving with slow, thermal velocities relative to the dark atom in the detector, interacts with the $X$He. The dark atom becomes polarized via the Stark effect induced by the electric field of the approaching nucleus itself, effectively forming a dipole. This interaction enables the nucleus to transition into a low-energy bound state within the potential well of the $X$He-nucleus system's effective potential. The energy released in this capture process corresponds to a photon whose energy equals the sum of the nucleus's initial kinetic energy and the binding energy in the potential well. This energy release mechanism is ultimately registered as the ionization signal observed in the NaI(Tl) detectors of the DAMA experiment.

\begin{figure}[!h]
\centering
\includegraphics[scale=0.5]{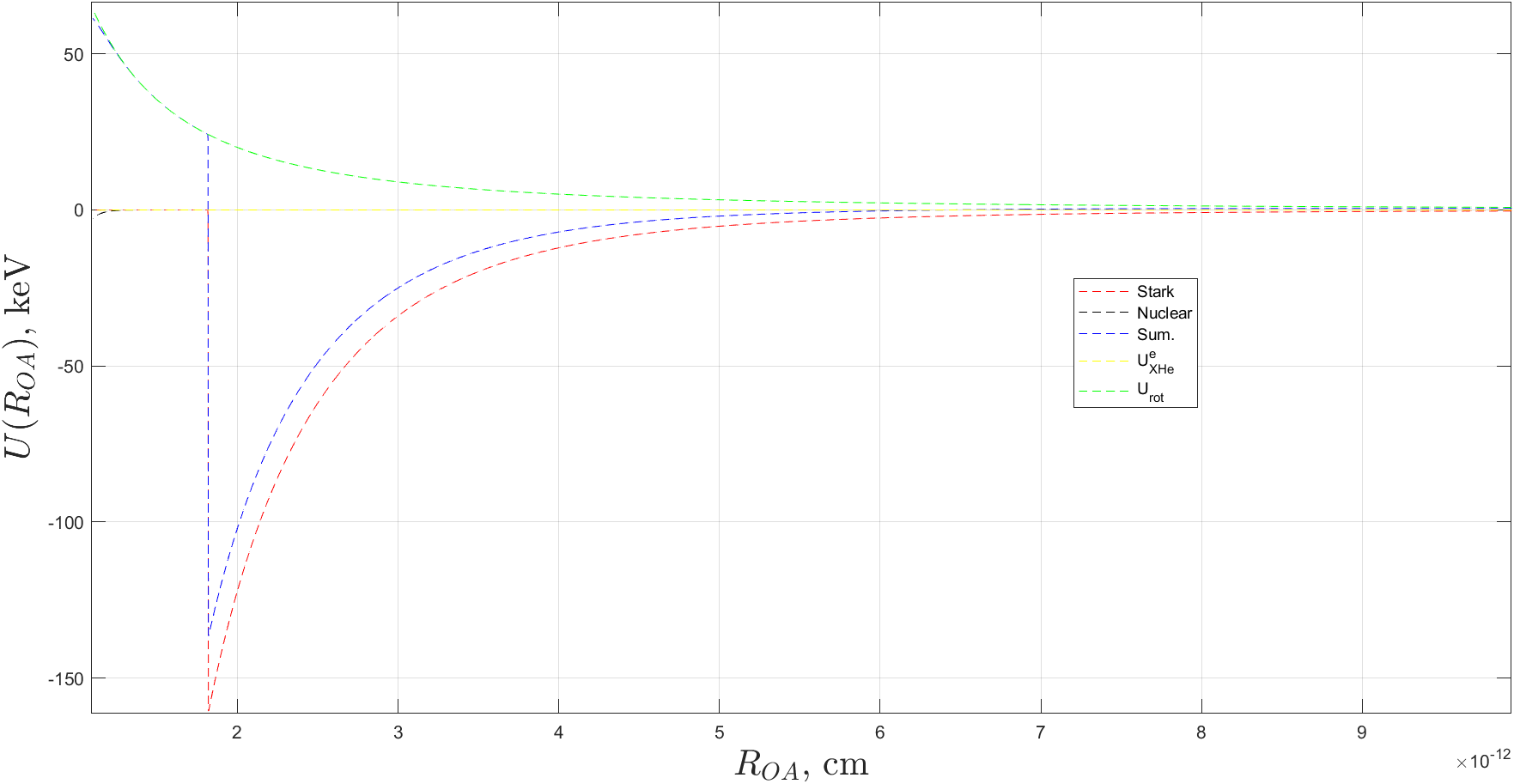}
\caption{
Interaction potentials within the OHe--Na system, presented as functions of the distance between the OHe dark atom and nucleus of Na, ${R}_{OA}$:
the Stark potential (red dotted curve), the centrifugal potential (green dotted curve), nuclear potential (black dotted curve), the electric interaction potential $U_{\rm XHe}^{\rm e}$ of unpolarized dark atom with the nucleus (yellow dotted curve) and the total effective interaction potential (blue dotted curve).
This particular configuration corresponds to a total angular momentum quantum number for the OHe-sodium nucleus interaction of $\Vec{J}_{\rm OHe-Na} = \overrightarrow{5/2}$. The calculations were performed utilizing the results of the \cite{Bikbaev_2025} paper.
}
\label{fig:Ris1}
\end{figure}

The theoretical description of interactions between dark atoms and ordinary nuclei constitutes a three-body problem, which is not amenable to an exact analytical solution. To elucidate the physical consequences of this scenario, characterized by its specific effective interaction potential, a precise quantum mechanical numerical model for this three-body system has been constructed. In article \cite{Bikbaev_2025}, a quantum mechanical numerical model describes the $O$He-$Na$ system, representing it as three particles interacting via electromagnetic, nuclear, and centrifugal forces. The computational approach involves solving the Schrödinger equation for the helium in the $O$He-$Na$ system at different fixed positions of the sodium nucleus, $\Vec{R}_{OA}$, relative to the dark atom.
This methodology, which incorporates both nuclear and electromagnetic interaction characteristics, enables precise determination of dark atom polarization through calculated dipole moments of the $O$He atoms. These distance-dependent dipole moments, varying with separation between the sodium nucleus and dark atom, facilitate reconstruction of the Stark potential - a crucial component in constructing the total effective interaction potential for the $O$He-$Na$ system. The total effective interaction potential is formulated as the sum of multiple contributions (see Figure \ref{fig:Ris1}): the Stark potential, centrifugal potential, nuclear potential, and the electric interaction potential $U_{\rm XHe}^{\rm e}$ of an unpolarized dark atom with the nucleus. The latter two potentials exhibit short-range character, diminishing exponentially with increasing particle separation. Thus, the model presented in \cite{Bikbaev_2025} therefore represents advancement toward a self-consistent quantum mechanical description of dark atoms with unshielded nuclear attraction interacting with usual matter nuclei.

Building upon the results presented in \cite{Bikbaev_2025}, we construct the total effective interaction potential for the $O$He--$Na$ system. This enables the determination of energy level for low-energy bound state between the $O$He dark atom and sodium nucleus, and facilitates computation of the corresponding capture reaction cross-section (see Figure \ref{fig:Ris1}). The form of this potential is influenced by the spin of the $\rm O^{--}$ particle, as the centrifugal component of the $O$He-nucleus total effective interaction potential depends on this spin magnitude. The specific value of the $\rm O^{--}$ spin represents a model-dependent quantity that varies with the underlying particle physics framework \cite{Khlopov_2020}. For the scenario depicted in Figure \ref{fig:Ris1}, the spin is taken as $I_{\rm O^{--}}=1$. The illustrated potential profile demonstrates consistency with theoretical expectations, exhibiting a potential well of approximately 136 keV depth preceded by a repulsive potential barrier. The height of this barrier significantly exceeds the thermal energy of sodium nuclei at room temperature (approximately $10^{-2}~\text{eV}$). This potential barrier plays a crucial role in maintaining dark atom integrity by preventing the merger of either constituent (helium or $\rm O^{--}$) with nucleus.

The bound states of sodium nucleus within the total effective interaction potential of the $O$He--$Na$ system (depicted by the blue dotted curve in Figure~\ref{fig:Ris1}) are obtained by solving one-dimensional stationary Schrödinger equation for free sodium nucleus in the corresponding potential. This procedure yields the discrete energy spectrum of bound states localized in the potential well, along with their associated normalized wave functions.

\begin{figure}[!h]
\centering
\includegraphics[scale=0.5]{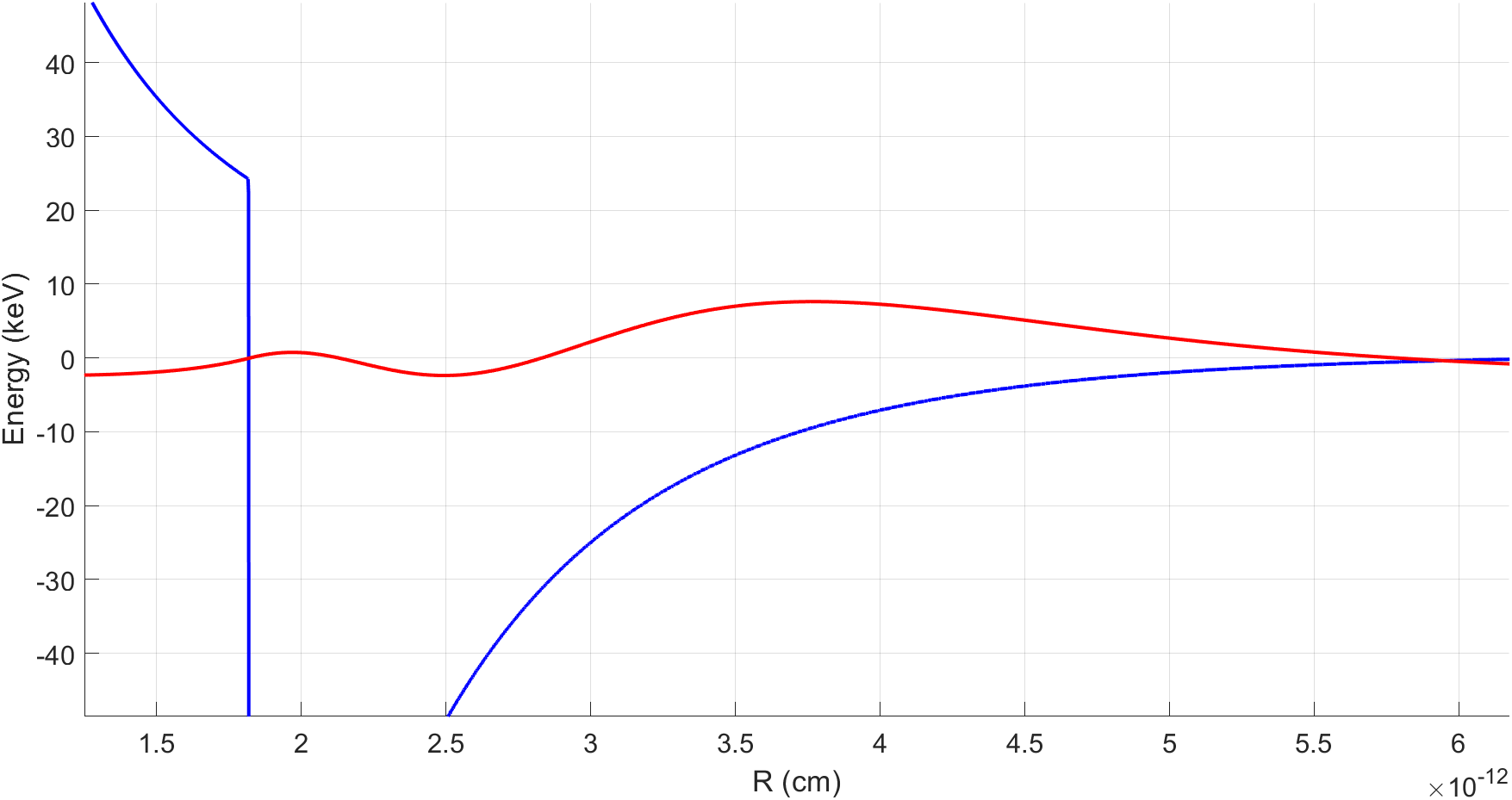}
\caption{
The figure illustrates the dependence of the total effective interaction potential between the $O$He dark atom and the sodium nucleus (solid blue curve) and the probability density given by the squared modulus of the wave function (solid red curve) from the radius vector of sodium nucleus. Squared modulus of the wave function correspond to the ground state energy level of $E_{1_{Na}}\approx-2.4\hspace{1mm}\text{keV}$ for sodium within the $O$He--$Na$ system's effective potential.
}
\label{fig:bind_energy_state_Na}
\end{figure}

The solution to this quantum mechanical problem is presented in Figure~\ref{fig:bind_energy_state_Na}. Analysis reveals that the potential well contains only a single bound state, corresponding to the ground state of the system with energy $E_{1_{Na}}\approx-2.4\hspace{1mm}\text{keV}$. The Figure~\ref{fig:bind_energy_state_Na} displays the total effective interaction potential (solid blue curve) alongside the squared modulus of the wave function  (solid red curve) for this single bound state within the $O$He--$Na$ potential.

\section{Calculation of the cross-section of radiative capture in the OHe-nucleus system}

We now proceed to compute the radiative capture cross-section for sodium nucleus into the bound state of the $O$He--$Na$ system. This calculation employs the previously derived unit-normalized wavefunction $\Psi_{f_{Na}}$, which describes the sodium nucleus in its ground bound state and will be consider as the final quantum state in the capture process.

In the initial configuration, the sodium nucleus exists as an unbound free particle represented by its wave function $\Psi_{i_{Na}}(r)$. This eigenfunction satisfies the Schrödinger equation governing the thermal relative motion of the nucleus of matter in the total effective interaction  potential $V_{eff}(r)$ characterizing the $O$He--$Na$ system:

\begin{equation}
\left[ -\frac{\hbar^2}{2\mu} \nabla^2 + V_{eff}(r) - E \right] \Psi_{i_{Na}}(r) = 0,
\label{eq_Shred_psii_2}
\end{equation}
where $\mu$ is the reduced mass of the $O$He-$Na$ system, $E=\cfrac{3}{2} k_{b}T$ is the energy of relative thermal motion in the center of mass system.

Owing to the spherical symmetry of the potential $V_{eff}(r)$, the wave function $\Psi_{i_{Na}}(r)$, which is modified by its influence, can be expressed as an expansion in partial waves corresponding to the angular momentum eigenfunctions:

\begin{equation}
\Psi_{i_{Na}}(r) = \sum_{l=0}^{\infty} \sum_{m=-l}^{l} \frac{u_l(r)}{r} Y_{lm}(\theta, \varphi),
\label{eq_psii_decomp_2}
\end{equation}
where $u_l(r)$ is the radial wave function, and $Y_{lm}(\theta, \varphi)$ these are spherical harmonics.

Inserting the partial wave expansion (\ref{eq_psii_decomp_2}) into the Schrödinger equation (\ref{eq_Shred_psii_2}) and applying the orthogonality relations of spherical harmonics yields the radial equation for individual partial waves:

\begin{equation}
\left[ -\frac{\hbar^2}{2\mu} \frac{d^2}{dr^2} + V_{eff}(r) + \frac{\hbar^2 l(l+1)}{2\mu r^2} - E \right] u_l(r) = 0,
\label{eq_Shred_u_l_2}
\end{equation}
where the term $\frac{\hbar^2 l(l+1)}{2\mu r^2}$ represents the centrifugal potential arising naturally from the separation of variables in spherical coordinates.

Thus, throughout the spatial domain where the effective interaction potential possesses non-zero values, the initial state wave function admits the following expansion:

\begin{equation}
\Psi_{i_{Na}}(r) = \sum_{l=0}^{\infty} (2l+1) i^l R_l^{\text{norm}}(r) P_l(\cos\theta),
\label{eq_Psi_i_Na_Norm}
\end{equation}
where $R_l^{\text{norm}}(r) =N_l \cdot u_l(r)/r=N_l \cdot R_l(r)$ is the normalized radial component of wave function, $N_l$ is the normalization factor, and $P_l(\cos\theta)$ denotes the Legendre polynomials.

The normalization of the radial wave function $R_l^{\text{norm}}(r)$ is chosen such that in the asymptotic limit ($r\to\infty$) it satisfies the condition:

\begin{equation}
R_l^{\text{norm}}(r) =  e^{i\delta_l}\left[ \cos\delta_l \, j_l(k_{Na}r) - \sin\delta_l \, n_l(k_{Na}r) \right],
\label{eq_R_l_asymptotics_2}
\end{equation}
where $k_{Na}=\cfrac{p_{Na}}{\hbar}=\cfrac{m_{Na}{\upsilon}}{\hbar}$ represents the wave number of the sodium nucleus, ${p}{Na}$ and $m{Na}$ denote its momentum and mass respectively, and ${\upsilon}$ is the relative velocity between interacting particles in the $O$He--$Na$ system. Here $j_l(kr)$ and $n_l(kr)$ correspond to spherical Bessel and Neumann functions, while $\delta_l$ represents the scattering phases determined by the effective potential $V_{eff}(r)$, which fully characterize the scattering behavior for each partial wave with orbital angular momentum $l$.

The scattering phases $\delta_l$ are determined through numerical integration of the radial Schrodinger equation (\ref{eq_Shred_u_l_2}) to obtain the wavefunction $R_l(r)$, which is subsequently matched to its asymptotic form given by equation (\ref{eq_R_l_asymptotics_2}). The phase shift computation employs the logarithmic derivative method, evaluating the quantity $L = R_l'(r_0)/R_l(r_0)$ at a sufficiently large radial coordinate $r_0$ where the effective potential $V_{eff}(r)$ becomes negligible:

\begin{equation}
\tan\delta_l = \frac{k_{Na} j_l'(k_{Na}r_0) - j_l(k_{Na}r_0) \cdot L}{k_{Na} n_l'(k_{Na}r_0) - n_l(k_{Na}r_0) \cdot L},
\end{equation}
where $j_l'$ and $n_l'$ represent the derivatives of the spherical Bessel and Neumann functions, respectively.

The effective interaction potential supports only one bound state for the sodium nucleus within the 1--6 keV energy range, restricting possible quantum transitions to an electric dipole (E1) process from an initial $l_i=1$ partial wave to the final $l_f=0$ bound state. Although thermal-energy nuclei predominantly occupy s-wave states ($l=0$), the initial state wavefunction contains a minor p-wave ($l=1$) admixture. We therefore employ a partial wave decomposition of the unbound free nuclear wavefunction, where each radial component satisfies the Schrödinger equation for the relative thermal energy $E=\frac{3}{2}k_{B}T$ in the $O$He--$Na$ center-of-mass system. Selecting the $l_i=1$ component from this expansion enables computation of the transition amplitude from this initial p-wave state to the final bound state. The reaction rate is significantly attenuated owing to the negligible population of p-wave states compared to the dominant s-wave component at thermal energies.

In accordance with Fermi's Golden Rule, the probability of transition per unit of time from an initial quantum state $\displaystyle |i\rangle$ to a specific final state $\displaystyle |f\rangle$ is given by:
\begin{equation}
\Gamma_{i\to f}={\frac{2\pi}{\hbar}}\left|\displaystyle\langle f|\hat{H}_{\text{int}}|i\rangle\right|^{2}g(E_{f}),
\end{equation}
where $\displaystyle\langle f|\hat{H}_{\text{int}}|i\rangle$ represents the transition matrix element of the interaction operator, $\hat{H}_{\text{int}}$, for the electrical transition between the final and initial states, and $\displaystyle g(E_{f})$ denotes the density of final states at energy $\displaystyle E_{f}$.

Fermi's golden rule expresses the transition rate between quantum states in terms of the density of available final states. For the capture process of a sodium nucleus by a dark atom accompanied by photon emission, this density of the final states is determined by the emitted photon. In the specific reaction under consideration, the sodium nucleus undergoes a transition from an unbound state to a bound configuration with the dark atom, emitting a photon whose energy is given by:
\begin{equation}
E_{\gamma}=T_{\rm{Na}}+I_{\rm{O}He-Na}\approx I_{\rm{O}He-Na}\approx 2\hspace{1mm}keV,
\label{eq_E_gamma_bound}
\end{equation}
where $T_{\rm{Na}}=\cfrac{p_{Na}^2}{2m_{Na}} \approx 10^{-2}~\text{eV}$ represents the thermal kinetic energy of the unbound free sodium nucleus, and $I_{\rm{OHe-Na}}$ denotes the binding energy of the sodium nucleus in the total effective interaction potential.

The initial configuration of the system consists of unbound sodium nucleus and $O$He dark atom, while the final state includes the bound $O$He--$Na$ system accompanied by photon emission. The continuum of final states is characterized by the photon parameters, as the initial sodium state exists in the continuous spectrum (free particle) while the final bound state is discrete. The transition becomes physically permissible only through photon emission, where the photon states themselves form a continuous spectrum. Consequently, the total density of final states is governed by the photon, since the $O$He--$Na$ bound state represents a fixed discrete configuration, whereas the photon can occupy various momentum and directional states.

The bound $O$He--$Na$ system possesses a discrete energy eigenvalue (approximately $2\hspace{1mm}\text{keV}$) following the sodium nucleus capture, thus its contribution to the density of final states  \(g(E_f)\) corresponds to a single quantum state, represented by a Dirac delta function. In contrast, the emitted photon, with energy closely matching the binding energy $I_{\rm{OHe-Na}}$, may be emitted in any spatial direction with essentially fixed energy (when neglecting the recoil of the $O$He--$Na$ system), which determines the angular distribution of the final state density. Consequently, the number of end states per unit energy interval and unit volume for photon emission into solid angle $d\Omega$, accounting for the two possible polarization states of the electromagnetic wave, in three-dimensional space is given by:
\begin{equation}
g(E_{\gamma}) =2 \cfrac{d}{dE_{\gamma}} \Biggl(\int \frac{d^3 \Vec{q}_{\gamma}}{(2\pi)^3} \Biggr)= \frac{E_{\gamma}^2 }{4\pi^3 c^3 \hbar^3}d\Omega,
\end{equation}
where $|\Vec{q}_{\gamma}|=\cfrac{E_{\gamma}}{c\hbar}$ denotes the photon wave vector.

The radiative capture cross section for sodium nucleus forming bound state with $O$He is defined by the relation:
\begin{equation}
\sigma_{\rm{OHe-Na}} = \frac{\Gamma_{i\to f}}{j},
\end{equation}
where $j$ represents the incident flux of sodium nuclei. 

For the radiative capture process, the initial state corresponds to scattering wave function normalized to unit flux, ensuring the probability flux associated with the incident wave satisfies:
\begin{equation}
j = \frac{\hbar k_{NA}}{\mu} = \upsilon
\end{equation}
where $\upsilon$ represents the relative velocity between the interacting particles. This relative velocity corresponds to the thermal velocity of the sodium nucleus relative to the dark atom in the center-of-mass system. For the specific capture process analyzed here, the substantial mass difference between the dark atom and sodium nucleus results in the center-of-mass system being effectively coincident with the laboratory system where the dark atom rests.

Substituting $ \Gamma_{i\to f} $ into the formula for the cross sections, we get:
\begin{equation}
\sigma_{\rm{OHe-Na}}={\frac{2\pi}{\hbar}}\cfrac{1}{\upsilon_{}}\left|\displaystyle \langle f|\hat{H}_{\text{int}}|i\rangle  \right|^{2}g(E_{\gamma})={\frac{2\pi}{\hbar}}\cfrac{1}{\upsilon_{}}\left|\displaystyle \langle f|\hat{H}_{\text{int}}|i\rangle  \right|^{2}\frac{E_{\gamma}^2 }{4\pi^3 c^3 \hbar^3}d\Omega.
\label{eq_sigma_OHe_Na_cross_sec}
\end{equation}

The transition of sodium nucleus into bound state with dark atom is mediated by its interaction with the electromagnetic field. The interaction Hamiltonian $\hat{H}_{\text{int}}$ governing the electric multipole transition of order $J$ in the dipole approximation is derived from the multipole expansion of the electromagnetic vector potential into functions with a certain moment and parity. For the long-wavelength approximation, which is fully applicable to the low-energy radiative capture process considered here given that $({q}_{\gamma}\cdot r)\approx 10^{-4}<<1$, the operator for electric multipole transition of order $J$ is given by the expression \cite{Davydov_1958}:

\begin{equation}
\hat{H}_{\text{int}} = -A_{0}\sqrt{\frac{2\pi(J+1)}{J[(2J+1)!!]^2}} q_{\gamma}^J \hat{Q}_{Jm},
\label{eq_H_int_OHe_Na}
\end{equation}
here, \(\hat{Q}_{Jm} = eZ_{Na} r^J Y_{Jm}(\theta,\phi)\) represents the static electric multipole moment operator, with $Z_{Na}$ denoting the charge number of the sodium nucleus, while $A_{0}=\sqrt{\cfrac{2\pi c\hbar}{q_{\gamma}}}$ corresponds to the electromagnetic vector potential amplitude, conventionally normalized to the single-photon per unit volume condition.

Consequently, the matrix element $\displaystyle\langle f|\hat{H}_{\text{int}}|i\rangle$ for the transition operator between the initial and final states can be evaluated. Employing the representation of the initial state wave function from Eq.~(\ref{eq_Psi_i_Na_Norm}) as a partial wave with definite orbital angular momentum $l{i}$, this matrix element factorizes into distinct radial and angular components as follows:

\begin{equation}
\langle f|\hat{H}_{\text{int}}|i\rangle = -A_{0}\sqrt{\frac{2\pi(J+1)}{J[(2J+1)!!]^2}} q_{\gamma}^J eZ_{Na}\langle l_f|Y_{Jm}(\theta,\phi)|l_i\rangle \cdot I_{\text{radial}},
\end{equation}
where \(I_{\text{radial}} =\int_0^\infty \Psi_{f_{Na}}^{*}(r)\cdot r^{J+2} \cdot (2l_{i}+1) i^{l_{i}} \cdot R_{l_{i}}^{\text{norm}}(r) dr\) and $\langle l_f|Y_{Jm}(\theta,\phi)|l_i\rangle$ are the radial and angular parts of the matrix element, respectively.

The angular component of the transition matrix element for the process \(l_i = J\to l_f = 0\) is determined through integration over the solid angle:
\begin{equation}
\begin{split}
\langle 0| Y_{Jm}(\theta,\phi)|J\rangle =\int Y_{00}(\theta,\phi)\cdot Y_{Jm}(\theta,\phi) \cdot P_{J}(cos(\theta)) d\Omega= \\ =\frac{1}{\sqrt{(2J+1)}}.
\end{split}
\end{equation}

Consequently, the squared modulus of the reduced matrix element for the static electric multipole moment operator $\hat{Q}{Jm}$ takes the form:
\begin{equation}
|\langle 0||\hat{Q}_{Jm}||J\rangle|^2 = \frac{e^2 Z_{Na}^2}{(2J+1)} \left| \int_0^\infty \Psi_{f_{Na}}^{*}(r)\cdot r^{J+2} \cdot (2J+1) i^{J} \cdot R_1^{\text{norm}}(r) dr \right|^2.
\end{equation}

Incorporating the expression of matrix element $\displaystyle\langle f|\hat{H}_{\text{int}}|i\rangle$ into the radiative capture cross-section formula (\ref{eq_sigma_OHe_Na_cross_sec}) for the $O$He--$Na$ bound state, and taking into consideration, that $e^2=\alpha\hbar c$, where $\alpha$ denotes the fine structure constant, we arrive at:

\begin{equation}
\begin{split}
\sigma_{\rm{OHe-Na}}^{J \to 0} = \frac{8\pi q_\gamma^{2J+1}\alpha c Z_{Na}^2}{\upsilon} \frac{(J+1)(2J+1)}{J[(2J+1)!!]^2}  \left| \int_0^\infty \Psi_{f_{Na}}^{*}(r)\cdot r^{J+2} \cdot i^{J} \cdot R_1^{\text{norm}}(r) dr \right|^2.
\end{split}
\label{eq_sigma_OHe_Na_cross_sec_2}
\end{equation}

The resulting formulation for the radiative capture cross section of sodium nucleus transitioning from unbound free state with orbital angular momentum $l_i = 1$ to the ground bound state (with $l_f = 0$) within the total effective interaction potential of the $O$He--$Na$ system and the corresponding the rate of radiation capture, is given by: 

\begin{equation}
\sigma_{\rm{OHe-Na}}^{1 \to 0} = \frac{16\pi}{3} \frac{\alpha c}{\upsilon} \cfrac{E_\gamma^3}{c^3\hbar^3} Z_{\text{Na}}^2 \left| \int_0^\infty \Psi_{f_{Na}}^{*}(r)\cdot r^{3} \cdot i \cdot R_1^{\text{norm}}(r) dr \right|^2,
\label{eq_sigma_OHe_Na_cross_sec_3}
\end{equation}

\begin{equation}
(\sigma_{\rm{OHe-Na}}^{1 \to 0}\cdot\upsilon)=\frac{16\pi}{3} \alpha c \cfrac{E_\gamma^3}{c^3\hbar^3} Z_{\text{Na}}^2 \left| \int_0^\infty \Psi_{f_{Na}}^{*}(r)\cdot r^{3} \cdot i \cdot R_1^{\text{norm}}(r) dr \right|^2.
\label{eq_sigma_OHe_Na_cross_sec_v_Na}
\end{equation}

Upon numerical evaluation of expressions (\ref{eq_sigma_OHe_Na_cross_sec_3}) and (\ref{eq_sigma_OHe_Na_cross_sec_v_Na}) using the relevant physical parameters -- including the matrix element squared \\ $\left| \int_0^\infty \Psi_{f_{Na}}^{*}(r)\cdot r^{3} \cdot i \cdot R_1^{\text{norm}}(r) dr \right|^2\approx1.8\cdot 10^{-65}\hspace{1mm}cm^5$, velocity ratio $\frac{\upsilon}{c}= \\ =\sqrt{\cfrac{3kT}{m_{Na}}}\approx2\cdot 10^{-6}$, nuclear charge $Z_{Na}=11$, and energy of the $O$He-$Na$ bound state $E_{\gamma}\approx-2.4\hspace{1mm}keV$ -- we obtain the following quantitative results for the radiative capture cross-section and corresponding radiative capture rate:

\begin{equation}
\sigma_{\rm{OHe-Na}}^{1 \to 0}\approx2.8\cdot 10^{-35}\hspace{1mm}cm^2=2.8\cdot 10^{-11}\hspace{1mm}barn.
\label{eq_sigma_OHe_Na_cross_sec_4}
\end{equation}

\begin{equation}
(\sigma_{\rm{OHe-Na}}^{1 \to 0}\cdot\upsilon)\approx1.6\cdot 10^{-30}\hspace{1mm}cm^3/sec.
\label{eq_sigma_OHe_Na_cross_sec_v_Na_2}
\end{equation}

The capture rate can be evaluated using the formalism presented in \cite{Khlopov:2010ik}:
\begin{equation}
\centering
    \begin{split}
        &R=1\cdot\cfrac{\rho_O}{M_O}(\left<\sigma v\right>_{Na}+\left<\sigma v\right>_{I})\cdot N_T,
        \\
        &\rho_O=\cfrac{M_O\,n_0}{320\cdot S_3\cdot30^{1/2}}V_h+\cfrac{M_O\,n_0}{640\cdot S_3\cdot30^{1/2}}V_E\cos(\omega(t-t_0)),
    \end{split}
\label{eq_capture_rate_substituted_parameters}
\end{equation}
where the scaling parameter $S_3=M_O/1~\text{TeV}$, with  the $O$He mass equal to $M_O=2.5~\text{TeV}$, the number of targets $N_T= 4.015\cdot 10^{24}$ nuclei per kilogram of NaI(Tl), Solar system velocity $V_h=220\cdot{10^5}\,\mbox{cm}/\mbox{s}$, Earth's orbital velocity $V_E=30\cdot{10^5}\,\mbox{cm}/\mbox{s}$, and local dark matter concentration $n_0=1.5\cdot10^{-4}\,\mbox{cm}^{-3}$ for the considered particles. The substantial mass of the DAMA/LIBRA detectors (approximately 9.7 kg each) ensures near-complete absorption of the low-energy gamma radiation inside the active detector volume.

Employing the equation (\ref{eq_capture_rate_substituted_parameters}), the capture rate can be computed utilizing the derived value for the rate of sodium radiative capture (\ref{eq_sigma_OHe_Na_cross_sec_v_Na_2}), under the assumption that the radiative capture rate for iodine is significantly suppressed:
\begin{equation}
    \begin{split}   
    R_{numerical}&\approx 0.440 + 2.83\times10^{-2}\cos(\omega(t - t_0))\, \mbox{counts/day kg},
    \end{split}
\end{equation}
this numerical result closely aligns with the experimental data reported by the DAMA/NaI and DAMA/LIBRA experiments.

Specifically, considering the energy interval from 1 keV to 6 keV, experimental analysis yields the following results:
\begin{equation}
    \begin{split}    \Delta R &= ( 6.95 \pm 0.45 )  \times 10^{-2}\, \mbox{counts/(day kg)},
    \\  
    R_0 &< 0.5 \, \mbox{counts/(day kg)},
    \end{split}
\label{eq_capture_rate_experimental_data}
\end{equation}
the modulated signal amplitude $\Delta R$ is determined by integrating the annual modulation amplitudes reported by the DAMA/NaI and DAMA/LIBRA collaborations \cite{BERNABEI_2020,BERNABEI_2021,BERNABEI_2022} over the energy range from the energy threshold to 6 keV. The upper limit $R_0$ on the unmodulated component is inferred from the corresponding constraints on the constant signal rate provided in the same references \cite{BERNABEI_2020,BERNABEI_2021,BERNABEI_2022}.

\section{Conclusions}

The developed quantum mechanical model provides a self-consistent description of the interaction between $O$He dark atom and atomic nucleus, addressing the fundamental challenges in direct dark matter detection. Through numerical solution of Schrödinger equations in the three-body system, we have reconstructed the total effective interaction potential for the $O$He--$Na$ system, revealing its characteristic form, comprising a shallow potential well and repulsive potential barrier. This potential configuration ensures the stability of dark atoms against nuclear fusion while permitting the formation of low-energy bound states through radiative capture processes.
The wave functions of the initial state of the free nucleus of matter, $\Psi_{i_{Na}}(r)$, and the final ground bound state of the nucleus of matter in the $O$He-$Na$ system, $\Psi_{f_{Na}}(r)$, are calculated by numerically solving the Schrodinger equation in the restored total effective interaction  potential of the considered system of three bodies. 
 
Our calculations demonstrate that the $O$He--$Na$ system supports exactly one bound state within the 1--6 keV energy range, corresponding to the ground state with binding energy $E_{1_{Na}}\approx-2.4$ keV. The radiative capture cross-section for the electric dipole transition from the initial $l_i=1$ partial wave to the final $l_f=0$ bound state yields $\sigma_{\rm{OHe-Na}}^{1 \to 0}\approx2.8\times10^{-35}$ cm$^2$, with the corresponding capture rate $\langle\sigma v\rangle\approx1.6\times10^{-30}$ cm$^3$/s.

The computed count rate of $R_{numerical}\approx 0.440 + 2.83\times10^{-2}\cos(\omega(t-t_0))$ counts/(day·kg) exhibits agreement with the annual modulation signal observed by DAMA/NaI and DAMA/LIBRA experiments in the 1--6 keV energy window where $\Delta R = (6.95 \pm 0.45)\times10^{-2}$ counts/(day·kg). This consistency provides substantial support for the $X$-helium dark atom hypothesis as a viable explanation for the DAMA results.

The methodology established in this work offers a robust foundation for further investigations into dark atom-nucleus interaction. Future research directions should take into account the not-point-like of interacting particles in the quantum mechanical model and include a detailed study of the dependence of capture rates for other conditions and detector materials.

\section*{Acknowledgements}
The work of T.B. was performed in Southern Federal University with financial support of grant of Russian Science Foundation № 25-07-IF. The work by A.M. was performed with the financial support provided by the Russian Ministry of Science and Higher Education, project “Fundamental and applied research of cosmic rays”, No.~FSWU-2023-0068. 



\end{document}